\documentclass[cits]{PoS}

\newcommand{\mean}[1]{\langle #1 \rangle} 

\title{``Jets'' and their distortion in heavy-ion collisions}

\ShortTitle{``Jets'' and their distortion in heavy-ion collisions}

\author{\speaker{Nicolas Borghini}\\
        Fakult\"at f\"ur Physik, Universit\"at Bielefeld, 
        Postfach 100131, D-33501 Bielefeld, Germany\\
        E-mail: \email{borghini@physik.uni-bielefeld.de}}

\abstract{After a discussion on the meaning of ``jets'' in the context of 
  nucleus--nucleus collisions, the distortions of the profile of a parton 
  shower induced by the presence of a medium is investigated in a QCD-inspired 
  model that implements the conservation of energy at each step of the shower 
  evolution.}

\FullConference{4th international workshop \ High-pT physics at LHC 09\\
		 February 4-7, 2009\\
		 Prague, Czech Republic}

\begin{document}

The large amount of data already registered in heavy-ion collisions at RHIC and 
the prospect of an extended kinematical reach at the LHC are shifting the focus
of studies of particles with high transverse momenta. 
While single-particle inclusive distributions of identified hadrons were at 
first the main topic of interest, attempts are now being made at evidencing 
structures involving several particles~\cite{Putschke:2008wn,Salur:2008hs,%
  Ploskon:QM2009}, using techniques~\cite{Cacciari:2008gn} primarily developed 
to identify jets in hadronic collisions. 
The purpose of these endeavors is to perform ``jet physics'' in nucleus--nucleus
collisions, measuring for instance the medium-induced modification of 
fragmentation functions of partons into hadrons.

This necessitates in parallel the development of theoretical and/or 
phenomenological models, which would help relate the measurements to properties 
of the medium created in the collisions of heavy ions at high energies. 
A step towards such a theoretical approach will be addressed in Section~\ref{%
  s:mmMLLA}, in which the possible distortions of a parton shower are studied 
within a QCD-inspired model. 
Before that, I shall stress in Section~\ref{s:discussion} the implicit ideas and
assumptions that underly the philosophy of studying ``jets'' in heavy-ion 
collisions, as well as some of the biases that enter their modeling.

\section{On ``jets'' in a medium}
\label{s:discussion}

Collisions of heavy nuclei at ultrarelativistic energies result in a high 
multiplicity of emitted particles. 
This trivial statement naturally means that trying to identify experimentally 
jet-like structure is a daunting task, since the ``underlying event''---which
to a very good extent can be reliably accounted for in proton-(anti)proton 
collisions~\cite{Cacciari:2008gn}---now contributes so much to the signal in 
the detectors, that even its fluctuations might mimick some local excess of 
energy-momentum flow.

The high-multiplicity environment also implies fundamental difficulties on the 
modeling side, some of which I wish to spell out here. 
In that view, let me first recall the salient features of the description of 
jets ``in vacuum'' (i.e., as produced in e$^+$e$^-$ or pp/p$\bar{\rm p}$ 
collisions) within the modified leading logarithmic approximation (MLLA) of 
perturbative QCD~\cite{Mueller:1982cq,Dokshitzer:1988bq}.
\medskip

Let $E$ and $\Theta_0$ be the jet energy and opening angle, and for any jet 
particle with energy $k_0$ and momentum $k_\perp$ transverse to the direction
of the energy flow (``jet axis'') let
\begin{equation}
\label{defs_l&y}
\ell\equiv\ln\frac{E}{k_0}=\ln\frac{1}{x}, \quad y\equiv\ln\frac{k_\perp}{Q_0},
\end{equation}
where $Q_0$ is some infrared cutoff parameter and $k_\perp\geq Q_0$.
To double logarithmic accuracy in $\ell$ and single logarithmic accuracy in 
$y$, the soft partons emitted in the jet evolution interfere destructively in 
some regions of phase space. 
This {\em color coherence\/} effect results in the simple interpretation of the 
parton shower evolution in terms of a probabilistic cascade in which the 
successive parton branchings, governed by the leading-order Altarelli--Parisi 
splitting functions, are independent and satisfy angular ordering~\cite{%
  Mueller:1982cq,Dokshitzer:1988bq}.
As was quickly realized~\cite{Marchesini:1983bm}, this probabilistic view allows
the implementation of jet evolution in a Monte-Carlo simulation.
Additionally, one can compute analytically within MLLA or even including 
next-to-MLLA corrections the inclusive longitudinal~\cite{Dokshitzer:1988bq} 
and transverse~\cite{Arleo:2007wn} distributions of hadrons within a jet---or 
rather, the parton distributions, which are then mapped onto hadrons using the 
local parton-hadron duality (LPHD) hypothesis---, leading to a satisfactory 
agreement with experimental data.

Last but not least, the nice description within (N)MLLA of the characteristics 
of jets measured in elementary-particle collisions ultimately relies on two
implicit yet essential premises:
\begin{itemize}
\item first, that one can define a jet in a way that is manageable and stable 
  both experimentally and theoretically~\cite{Huth:1990mi,Ellis:1991kq}; 
\item second, that the object thus specified can be ``isolated'' from its 
  environment---be it the underlying event in a hadronic collision, or the 
  other jet(s) in events with at least two jets---, so that the properties 
  which are attributed to the jet are actually intrinsic to it. 
  This requirement restricts the range of characteristics that can meaningfully 
  be assigned to a jet~\cite{Dokshitzer:1988bq}.
\end{itemize}
\medskip

Turning now to the case of heavy-ion collisions, one can question which of the 
above features survive.

The assumption that there exists an object (the ``jet''), with definite 
characteristics that depend on the properties of the medium created in the 
nucleus-nucleus collision, yet which can be investigated theoretically as an 
independent entity, lies at the core of ``jet physics'' studies performed in
heavy-ion collisions, including that reported in Section~\ref{s:mmMLLA}.

It is also tacitly assumed that this ``jet'' can be, at least in thought, 
isolated from a ``background event'', so that one can meaningfully study their
respective influences on each other: 
on the one hand, the perturbation of the medium induced by the propagation of 
the jet---possibilities include shock-waves along a Mach cone~\cite{%
  Baumgardt:1975qv,CasalderreySolana:2004qm} or Cherenkov-like radiation of 
gluons~\cite{Dremin:2005an,Ruppert:2005uz,Koch:2005sx}---;
on the other hand, the distortions of the ``jet'' by the medium through which 
it propagates.
The picture underlying the last point implies that the adopted definition of a 
``jet'' also makes sense in the absence of a medium, i.e.\ in collisions of 
elementary particles, so that one has a reference with respect to which the 
medium-induced modifications can be investigated.

There very probably exists a satisfactory definition of a ``jet'', and quite 
possibly several ones. 
This belief is driven by the fact that the modification of the single-particle 
inclusive distribution, which is a clear observable, at high transverse momentum
by the medium-enhanced radiation of soft gluons~\cite{Baier:2000mf,%
  Gyulassy:2003mc,Kovner:2003zj} is a theoretically well-defined phenomenon.

\noindent The looked-for definition of a ``jet'' would however rather imply some
object involving more than one particle, with transverse momenta quite probably 
above a given lower cutoff.
The answer should definitely be driven by experimental attempts, by the 
practical feasibility. 
Meanwhile, theorists and phenomenologists can only speculate about the modeling 
of the ``jet'' in heavy-ion collisions.
\medskip

Given the expected properties of the ``background event'', namely that it 
reflects a medium which has a finite size with a non-trivial geometry and which 
is rapidly expanding, so that its properties change in space and time, the most 
natural approach is to look for a description of the ``jet'' which can be 
implemented in a Monte Carlo simulation~\cite{Zapp:2008gi,Armesto:2008qh}.
One could then embed the simulation of the ``jet'' into that of the ``background
event''~\cite{Renk:2008pp,Lokhtin:2008xi}, to compare the result with the 
experimental data. 

This requirement of a Monte Carlo approach is not assumption-free. 
If the ``jet'' is a multi-hadron object, then this involves almost automatically
the occurrence of branchings in its evolution.
From the technical point of view, a consistent probabilistic implementation 
necessitates that the branchings are {\em independent\/} from each other. 
For in-vacuum jets, this independence follows from the effects of color 
coherence. 
Yet whether this coherence is retained within a colored medium is far from 
obvious. 
Moreover, the physics of the enhanced radiation of gluons by a high-momentum 
parton in a colored medium is precisely that of a coherent emission {\`a} la
Landau--Pomeranchuk--Migdal (LPM)~\cite{Baier:2000mf}, at least for part of the 
emitted gluon spectrum.
It was recently argued that a probabilistic implementation of the LPM effect, 
relying on formation time arguments, can be found~\cite{Zapp:2008af}.
An alternate approach is the formulation of the LPM effect as a modification of 
parton splitting functions~\cite{Arnold:2008zu}.
These are encouraging results, which however might not be applicable to the 
whole phase space of produced gluons. 
Additionally, if color coherence in the ``jet'' development is not retained in 
a colored medium, one also loses the angular ordering of successive branchings,
which means that the ``jet'' structure becomes more complicated. 

Eventually, the choice of the physics ingredient which is put forward in the 
definition of the ``jet'' is not neutral either. 
The picture of the medium-induced modifications of a ``jet'' that has been 
adopted till now~\cite{Salgado:2003rv,Borghini:2005em} mostly emphasizes the 
conservation of energy-momentum.
That is, the ``jet'' is seen as an object inside which, through the influence of
the medium, energy and momentum are redistributed, so that the meaningful 
comparison would be between jets in vacuum and in heavy-ion collisions with 
the same energy.
The choice is greatly motivated by considerations on the (parton) production 
cross-sections, yet it should be kept in mind that there is some arbitrariness 
in it.
In particular, the exchange of energy and/or momentum of the ``jet'' with the 
``background event''---either the loss of energy by the jet through elastic 
scatterings in the medium~\cite{Bjorken:1982tu}, or the gain in mean-square 
transverse momentum due multiple collisions on the medium scattering 
centers~\cite{Baier:1996sk}---complicates the separation between both components
of the event if one really insists on conserving energy and momentum inside the 
``jet''.

\section{A ``medium-modified parton shower'' approach}
\label{s:mmMLLA}

Under all the caveats on the validity of the approach which were recalled in 
Section~\ref{s:discussion}, let me introduce my model of a ``medium-modified 
parton shower'' and some of its possible consequences~\cite{Borghini:2005em,%
  Borghini:2009eq}.

The idea is to mimick the medium-enhanced radiation of gluons by a fast parton 
by modifying the $q\to qg$ and $g\to gg$ splitting functions, enhancing their 
singular parts. 
To allow analytical computations, this enhancement is obtained by a 
multiplication of the singular parts by a constant factor $1+f_{\rm med}$. 
That is, energy is conserved at each branching, while the splitting probability 
is enhanced: one modifies the structure of the cascade, considered as an 
isolated entity; in particular, there is no transfer of momentum between the 
medium and the ``jet''. 

Under this setup, one can compute the longitudinal and transverse distributions 
of partons inside the resulting shower, repeating the steps that turned out to 
be successful in the description of in-vacuum jets. 
Note that imposing the conservation of energy while increasing the splitting 
probability automatically results in an increase of the ``jet'' 
multiplicity~\cite{Dremin:2006da,Armesto:2008qe,Ramos:2008qb}.

\subsection{Modification of the longitudinal distribution of partons}
\label{s:hump-backed-plateau}

The splitting functions enter evolution equations for the distributions 
${\cal D}_{\!A_0}^A(x,E\Theta,Q_0)$ of partons of type $A$ with the energy 
fraction $x$ inside a shower with opening angle $\Theta$ initiated by parton 
$A_0$ with energy $E$~\cite{Dokshitzer:1988bq}.
These equations (or rather, an approximation thereof) can be solved, leading in 
the limit $Q_0=\Lambda_{\rm QCD}$ to the {\em limiting spectrum\/} 
$\tilde{\cal D}^{\lim}(\ell,y)=x{\cal D}_g^{\rm MLLA}(\ell,y)$ with its 
characteristic ``hump-backed plateau'' shape~\cite{Dokshitzer:1988bq}.
Note that $y\approx\ln(k_0\Theta/Q_0)$, see Eq.~(\ref{defs_l&y}), and that the 
``partonic'' limiting spectrum is mapped through LPHD on the hadronic one 
$K^h\tilde{\cal D}^{\lim}$.
\medskip

\begin{figure}[b]
  \centerline{\includegraphics*[width=0.667\linewidth]{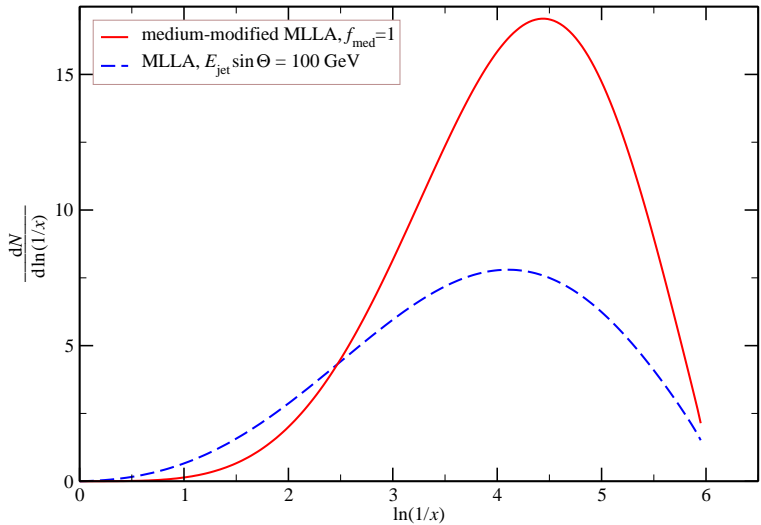}}
  \vspace{-2mm}
  \caption{\label{fig:longitudinal_spectrum}Longitudinal distribution inside a 
    parton shower initiated by a parton such that 
    $\displaystyle Y_\Theta\equiv\ln\frac{E\sin\Theta}{Q_0}=6$}
\end{figure}

The effect on the $\ell$-distribution of modifying the Altarelli--Parisi 
splitting functions through the introduction of the $1+f_{\rm med}$ factor has 
been studied in Ref.~\cite{Borghini:2005em}, and consists in redistributing 
partons from low-$\ell$ values to larger ones, as is illustrated in the case 
of a 100~GeV jet in Fig.~\ref{fig:longitudinal_spectrum}.
This follows at once from the assumed ingredients of the model: if the splitting
probability increases, this decreases the probability that partons with high 
energy fractions $x$ survive the evolution.
Since energy is to a large extent conserved, this must in turn result in a 
significant growth of the multiplicity at low $x$, i.e. large $\ell$.
The advantage of the model is that the redistribution is encoded in the 
analytical formula that gives the distorted longitudinal spectrum, which can 
then be used for other purposes.

\subsection{Transverse momentum distribution}
\label{s:kTdistribution}

The calculation of the distribution in transverse momentum with respect to the 
jet axis for hadrons inside a jet relies on integrating over one hadron ($h_2$) 
the double differential two-particle cross-section, weighted by the hadron 
energy fraction $x_2$~\cite{Dokshitzer:1988bq}. 
The process under consideration, which is represented e.g. in Fig.~1 of 
Ref.~\cite{PerezRamos:2005nh}, is the following.
A hard process gives rise to an ``initial'' parton $A_0$ with the typical energy
scale $E\Theta_0$.
This parton emits with the probability $D_{\!A_0}^A\!(u,E\Theta_0,uE\Theta)$, 
which is governed by the splitting functions, a parton $A$ that carries a 
fraction $u$ of the energy of the initial parton and has the virtuality 
$uE\Theta<E\Theta_0$.
Parton $A$ can then further evolve and split into two partons $B$ (energy $uzE$)
and $C$ (energy $u(1-z)E$) forming an angle of typical order $\Theta$; 
the corresponding probability again involves the parton splitting function 
$P_{BA}(z)$.
Eventually, the partons hadronize: $B$ gives some hadron $h_1$ with the energy 
$x_1E$, $C$ gives $h_2$ with the energy $x_2E$.
The probability for these hadronization processes $A\to h_i$ is given by the 
fragmentation function $D_{\!A}^{h_i}(x_i/u,uE\Theta,Q_0)$.
All in all, the distribution $F_{\!A_0}^h\!(x,\Theta,E,\Theta_0)$ of the energy 
fraction of hadrons within a subjet with an opening angle $\Theta<\Theta_0$ 
reads~\cite{Dokshitzer:1988bq}
\begin{equation}
F_{\!A_0}^h(x,\Theta,E,\Theta_0)\equiv 
\sum_{A=q,g} \int_x^1\!{\rm d}u\,D_{\!A_0}^A\!(u,E\Theta_0,uE\Theta)\,
   D_{\!A}^h\Big(\frac{x}{u},uE\Theta,Q_0\Big).\label{F_A0^h}
\end{equation}

Differentiating this distribution with respect to to $\ln\Theta$---or 
equivalently, at fixed $\ell$, $\ln k_\perp$---yields the double differential 
single-particle inclusive distribution of hadrons inside a jet initiated by the 
parton $A_0$ with opening angle $\Theta_0$ and energy $E$~\cite{%
  Dokshitzer:1988bq}:
\begin{equation}
\label{2-diff_dist}
\frac{{\rm d}^2N}{{\rm d}x\,{\rm d}\ln k_\perp} \simeq
\frac{{\rm d}^2N}{{\rm d}x\,{\rm d}\ln\Theta} =
\frac{\rm d}{{\rm d}\ln\Theta} F_{\!A_0}^h\!(x,\Theta,E,\Theta_0).
\end{equation}
Within a small-$x$ approximation and MLLA, this can be rewritten as~\cite{%
  PerezRamos:2005nh}
\begin{equation}
\label{2-diff_dist-bis}
\bigg(\frac{{\rm d}^2N}{{\rm d}\ell\,{\rm d}y}\bigg)_{\!\!g,q}\! =
\frac{\rm d}{{\rm d}y}
  \bigg[\frac{\mean{C}_{g,q}}{N_c}K^h\tilde{\cal D}^{\lim}(\ell,y)\bigg],
\vspace{-2mm}
\end{equation}
where $K^h\tilde{\cal D}^{\lim}$ is the limiting spectrum for hadrons (see above)
while the average color current $\mean{C}_{A_0}$ inside a jet initiated by $A_0$ 
follows from the splitting functions.
If one wants the distribution of partons, instead of hadrons, one should just 
divide by the constant $K^h$. 

The transverse momentum distribution inside a jet is finally given by the 
integral of Eq.~(\ref{2-diff_dist-bis}) over $\ell$:
\begin{equation}
\label{kTdist}
\bigg(\frac{{\rm d}N}{{\rm d}\ln k_\perp}\bigg)_{\!\!g,q}\! =
  \int\!{\rm d}\ell\,
  \bigg(\frac{{\rm d}^2N}{{\rm d}\ell\,{\rm d}\ln k_\perp}\bigg)_{\!\!g,q}.
\end{equation}
\medskip

The various steps in the computation can be repeated using ``medium-modified'' 
splitting functions instead of the Altarelli--Parisi functions.
Let me present the results, for gluon-initiated parton showers with the energy 
100 GeV.
\begin{figure}[h]
  \centerline{\includegraphics*[width=0.667\linewidth]{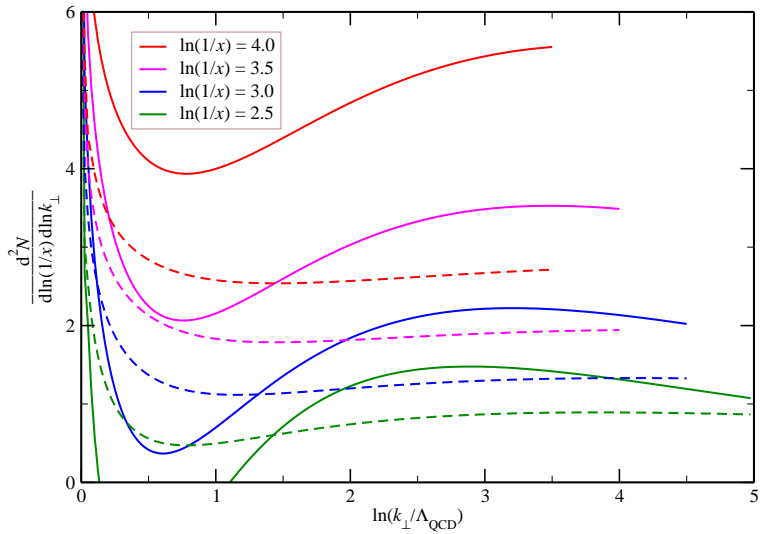}}
  \vspace{-2mm}
  \caption{\label{fig:doublediff_spectrum}Double differential distribution 
    inside a parton shower initiated by a gluon such that $Y_\Theta=6$, for 
    various values of $\ell$. 
    Dashed lines: MLLA; 
    full lines: ``medium-modified MLLA'' with $f_{\rm med}=1$.}
\end{figure}
Figure~\ref{fig:doublediff_spectrum} displays, for various values of the energy 
fraction $x$, the double differential single-particle distribution~(\ref{%
  2-diff_dist-bis}).
The dashed curves, representing in-vacuum jets, reproduce the calculations of 
Ref.~\cite{PerezRamos:2005nh}: at fixed $\ell$, the distribution is rather flat
for $y\gtrsim 1$, while the divergence as $y\to 0$ reflects that of the running 
QCD coupling constant $\alpha_s(k_\perp)$ when $k_\perp\to\Lambda_{\rm QCD}$, and 
thus signals the breakdown of the perturbative regime and thereby the validity 
of the computation.

\noindent The effect of increasing the singular parts of the splitting function 
is to deplete the low-$k_\perp$ region, while increasing the high-$k_\perp$ one.
The effect seems to be proportionally more marked for small $\ell$ values than 
at low $x$. 
Actually, for the lowest $\ell$ value reported here, the ``influence of the 
medium'' gives negative values to the particle distribution for 
$k_\perp\lesssim 3\Lambda_{\rm QCD}$. 
This is slightly disturbing, although obviously in a regime where even the 
vacuum calculation is not to be taken too literally. 
One should therefore only see the general qualitative trend of redistributing 
partons from low to large transverse momenta, without attaching too much value 
to the quantitative aspects.

\begin{figure}[b]
  \centerline{\includegraphics*[width=0.667\linewidth]{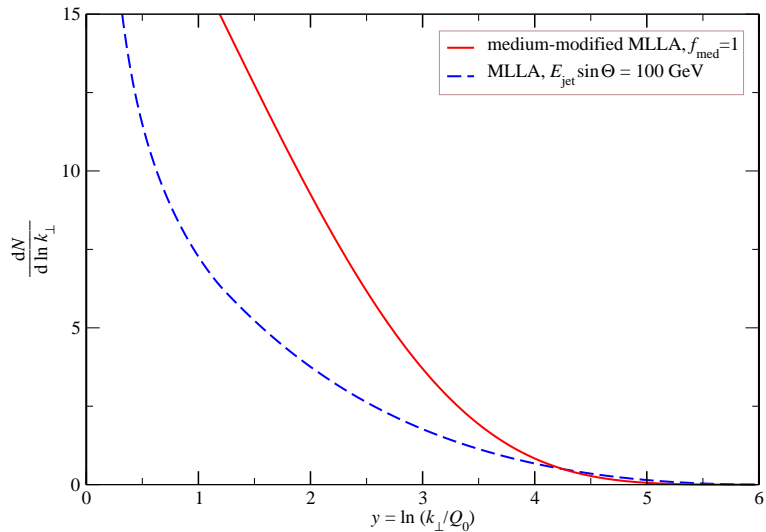}}
  \vspace{-2mm}
  \caption{\label{fig:transverse_spectrum}Transverse distribution inside a 
    parton shower initiated by a gluon such that $Y_\Theta=6$, as a function of 
    $\ln k_\perp$.}
\end{figure}

When integrating over $\ell$, this qualitative reshuffling from low to large 
$k_\perp$ values results in a broader $k_\perp$ range where the distribution is 
sizable, as shown on Fig.~\ref{fig:transverse_spectrum}, which shows the 
transverse momentum distribution~(\ref{kTdist}) as a function of $\ln k_\perp$.
The jet is thus ``broadened'' in transverse momentum.
One can similarly show that the same model leads to in-medium {\em angular 
  broadening\/}, the distribution in the angle $\Theta$ with respect to the jet 
axis being broader for the medium-distorted parton shower as in the 
vacuum~\cite{Borghini:2009eq}. 

\begin{figure}
  \centerline{\includegraphics*[width=0.667\linewidth]{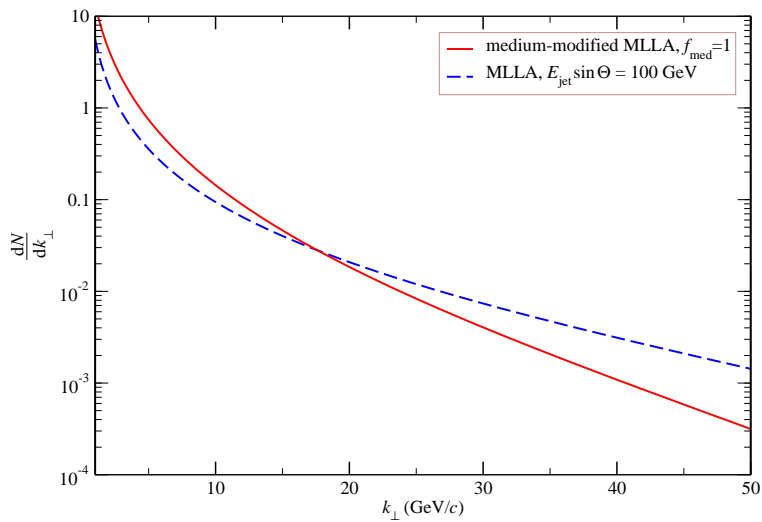}}
  \vspace{-2mm}
  \caption{\label{fig:transverse_spectrum2}Transverse distribution inside a 
    parton shower initiated by a gluon such that $Y_\Theta=6$, shown vs.\ 
    $k_\perp$.}
\end{figure}

At the same time however, one can guess on Fig.~\ref{fig:transverse_spectrum} 
that at large values of $k_\perp$ the distribution is actually smaller for 
in-medium ``jets'' as in the vacuum. 
This actually reflects the redistribution of partons from high to low $x$ 
values.
The combination of both effects is indeed the softening of the transverse 
momentum spectrum, as shown in Fig.~\ref{fig:transverse_spectrum2}.
This softening was previously also observed in simulations with the 
Q-{\sc Pythia} parton cascade~\cite{Armesto:2008qe}.

Note that a proper accounting in analytical calculations of the large transverse
momentum region might necessitate the introduction of next-to-MLLA corrections 
that keep track of energy conservation in the branchings with better 
accuracy~\cite{Arleo:2007wn}.
This is however at this time a rather academic point, given the exploratory 
character of the model. 
In particular, one can quite safely anticipate that the only experimentally 
accessible modifications of jet shapes have to be investigated above some lower 
cutoff in the momentum transverse to the beam axis, as was partly done in 
Refs.~\cite{Borghini:2005em,Borghini:2009eq}, but has to be studied in a more 
systematic way.

\end{document}